\documentclass[aps,twocolumn,prb,amsfonts,showpacs,superscriptaddress,longbibliography]{revtex4-2}
\usepackage{graphicx}
\usepackage{bm}
\usepackage{hyperref}
\usepackage{amsmath}
\usepackage{amssymb}
\usepackage[table]{xcolor}
\usepackage{array}
\usepackage{multirow}
\usepackage[caption=false]{subfig}
\usepackage{physics}
\usepackage{booktabs}
\usepackage{soul}
\usepackage{amsthm}
\usepackage{bbm}
\usepackage{multirow}
\usepackage[colorinlistoftodos]{todonotes}
\usepackage[scr=boondoxo,frak=boondox,scrscaled=1.05]{mathalfa}
\usepackage{amsmath}

\def\r{{\boldsymbol{r}}}
\def\f{{\boldsymbol{f}}}

\def\k{{\boldsymbol{k}}}

\def\v{{\boldsymbol{v}}}
\def\q{{\boldsymbol{q}}}
\def\A{{\boldsymbol{A}}}
\def\g{{\boldsymbol{g}}}

\def\R{{\boldsymbol{R}}}
\def\G{{\boldsymbol{G}}}

\def\b{{\boldsymbol{b}}}
\def\K{{\boldsymbol{K}}}
\def\L{{\boldsymbol{L}}}
\def\h{{\boldsymbol{h}}}
\def\w{{\boldsymbol{w}}}

\usepackage{ulem}
\begin{document}

\title{Topological Order in Neural Wavefunctions}

\author{Ahmed Abouelkomsan}
 \thanks{ahmed95@mit.edu}
\affiliation{Department of Physics, Massachusetts Institute of Technology, Cambridge, MA-02139, USA}

\author{Max Geier}

\affiliation{Department of Physics, Massachusetts Institute of Technology, Cambridge, MA-02139, USA}

\author{Liang Fu}
 \thanks{liangfu@mit.edu}
\affiliation{Department of Physics, Massachusetts Institute of Technology, Cambridge, MA-02139, USA}

\begin{abstract}

Topologically ordered states are among the most interesting quantum phases of matter that host emergent quasi-particles having fractional charge and obeying fractional quantum statistics. Theoretical study of such states is however challenging owing to their strong-coupling nature that prevents conventional mean-field treatment. Here, we demonstrate that an attention-based deep neural network provides an expressive variational wavefunction that discovers fractional Chern insulator ground states purely through energy minimization without prior knowledge and achieves remarkable accuracy. We introduce an efficient method to extract ground state topological degeneracy --- a hallmark of topological order --- from a single optimized real-space wavefunction in translation-invariant systems by decomposing it into different many-body momentum sectors. Our results establish neural network variational Monte Carlo as a versatile tool for discovering strongly correlated topological phases. 

\end{abstract}
\maketitle
\date{\today}

\section{Introduction}
The recent discovery of the fractional quantum anomalous Hall effect \cite{park2023observation,cai2023signatures,zeng2023thermodynamic,xu2023observation,lu2024fractional} opened an exciting frontier for studying fractional topological phases in quantum materials characterized by the emergence of fractional charge and fractional statistics. 
In both twisted MoTe$_2$ and rhombohedral graphene moir\'e systems, the observed fractional quantum anomalous Hall states 
feature a series of quantized Hall conductance plateaus in the absence of an external magnetic field, which exactly match those of fractional quantum Hall states in two-dimensional electron systems under strong magnetic fields. 
This striking correspondence can be understood from topological order \cite{wen2004quantum}---a unified description of fractionally quantized Hall conductance, fractional charge, and fractional quantum statistics.

Historically, topological order has long played a key role in the study of fractional Hall states in Landau levels and Chern bands.  
 As a matter of fact, prior to the recent experimental breakthrough, fractional quantum anomalous Hall states---or zero-field fractional Chern insulators \cite{tang2011high,neupert2011fractional,regnault2011fractional,sheng2011fractional} ---were theoretically predicted in twisted bilayer graphene \cite{abouelkomsan2020particle,repellin2020chern,ledwith_fractional_2020} and twisted transition metal dichalcogenides (TMDs)  \cite{li2021spontaneous,crepel2023anomalous} based on  the numerical diagnosis of topological orders. On the theory side, most computational studies have primarily relied on exact diagonalization (ED) and the density-matrix renormalization group (DMRG). In ED, topological order is typically diagnosed through ground-state degeneracy on the torus, the many-body Chern number, or entanglement spectroscopy.





While both methods have been rather successful, they suffer from notable drawbacks. ED, though unbiased, is limited to small system sizes and often restricted to few energy bands due to the exponential growth of the Hilbert space, and therefore cannot capture strong band-mixing effects \cite{abouelkomsan2024band,yu2024fractional}. On the other hand, DMRG can treat larger systems but is essentially a quasi-one-dimensional method, making it difficult to apply to strictly two-dimensional geometries. 


Recently, neural network variational Monte Carlo (NN- VMC) \cite{carleo2017solving} has emerged as a promising numerical method for strongly interacting systems which is able to overcome these drawbacks. First, VMC can be directly formulated in terms of first-quantized wavefunctions  making it well-suited for studying various systems such as moir\'e materials which are described by continuum models. 

Second, the computational complexity of VMC scales polynomially with the number of particles \cite{foulkes2001quantum}, implying that simulations can, in principle, be pushed to larger systems. In addition, the expressivity of neural networks enables a variational ansatz capable of describing highly nontrivial quantum phases, including topological order and chiral superconductivity \cite{teng2025solving,li2025deep,qian2025describing,luo2025solving,nazaryan2025artificial,li2025attention}. Importantly, as NN-VMC is carried out directly in real space, it accounts for \textit{all} energy bands without truncation, thereby capturing arbitrary band-mixing effects. Moreover, it readily adapts to various two-dimensional geometries.

However, the application of VMC to topological order has been challenging. In the VMC method, a trial many-body wavefunction $\Psi_{\{\theta\}}(\{\r_i\}) $ 
is optimized to minimize the ground state energy which raises a natural question, how to detect topological order from a single variational ansatz $\Psi_{\{\theta\}}(\{\r_i\}) $ ? As topological order is highly non-local, it is impossible to be diagnosed with any local observable. On one hand, topological order can be generally detected from a single wavefunction by extracting the topological part of entanglement entropy \cite{kitaev2006topological,levin2006detecting}. However, the approach has been proven to be quite delicate, requiring careful subtraction of area-law contributions and  the application of certain tricks to efficiently sample the reduced density matrix \cite{hastings2010measuring,hibat2020recurrent,hibat2023investigating,wang2020calculating}. It is not clear a priori if there are efficient ways to diagnose topological order in variational wavefunctions that possess a large number of parameters, e.g., weights and biases in neural wavefunctions. 

A hallmark of topological order is ground state degeneracy on closed-surface manifolds with non-zero genus $g\neq 0$ \cite{wen1990ground, niu1985quantized}. Unlike symmetry breaking states, topologically degenerate ground states are locally indistinguishable and only differ from each other in nonlocal observables. The existence of topological degeneracy is directly related to the existence of fractional quasi-particles, or anyons. In particular, topological degeneracy on the torus ($g=1$) equals the number of anyon types.   

In this work, we present an efficient diagnostic of topological order that infers ground-state topological degeneracy from a single many-body wavefunction. Our method applies to translation-invariant systems, where degenerate ground states are labeled by many-body (crystal) momentum. We introduce a post-processing protocol that decomposes one optimized variational wavefunction into distinct momentum sectors, thereby producing a set of (nearly) degenerate ground states. In essence, this “momentum spectroscopy” yields momentum-resolved variational wavefunctions—even though the neural network architecture and the VMC optimization are entirely momentum-agnostic.

This method is ideally suited for uncovering topologically degenerate ground states that occur at distinct momenta sectors.     
As a concrete example, we introduce and study a continuum model of interacting fermions in a periodic (emergent) magnetic field with zero net flux per unit cell. 
Due to the vanishing flux, it is unclear if topological ordered states such as fractional Chern insulators can exist. By training a general-purpose fermionic neural network through energy minimization, we find that the optimized neural wavefunction at $\nu=1/3$ filling corresponds to a gapped quantum liquid that appears to be featureless.  
Nonetheless, by applying our post-processing protocol, we find a clear three-fold topological degeneracy.  Additionally,  
the \textit{same} neural network finds the competing charge density wave (CDW) in our model under a different parameter regime.
 


\section{Momentum spectroscopy} We consider a quantum system on a finite torus described by a Hamiltonian $H$. Let $\Psi(\{\r_i\}) \equiv \Psi(\r_1 , \dots, \r_N)$ be a variational approximation to the many-body ground state of $N$ particles representing a topologically ordered phase. Our goal is to reveal the underlying topological degeneracy of the system from the information encoded in the \textit{single} wavefunction $\Psi(\{\r_i\})$. We assume that the system of interest has translational symmetry; $[H,t_i(\R)] = 0$ where $t_{i}(\R)$ is translational operator acting on the $i$-th particle translating it by a distance $\R$, $t_{i}(\R) \Psi(\r_1 , \dots, \r_N)\  = \Psi(\r_1, \cdots, \r_i + \R, \cdots, \r_N)$

Next,  we can expand $\Psi(\{\r_i\})$ in terms of a basis set of states which are eigenstates of the center of mass (COM) translation operator $T(\R) = \prod_i^N t_i(\R)$, 
\begin{equation}
\label{eq:decomposition}
    \Psi(\r_1, \dots, \r_N) = \sum_{\K} c_{\K} \Phi_{\K}(\{\r_i\}) 
\end{equation}
where \begin{equation}
\label{eq:projection}
    \Phi_{\K}(\{\r_i\}) = \frac{1}{N_s} \sum_{\R} e^{-i \K \cdot \R} \Psi(\r_1 + \R, \cdots, \r_N + \R)
\end{equation}

denote eigenstates of $T(\R)$, $T(\R) \Phi_{\K}(\{\r_i\}) = e^{i\K \cdot \mathbf{R}} \Phi_{\K}(\{\r_i\})$  with COM momentum $\K$ and $N_s$ is the number of distinct translations $\R$ that span the whole torus. Note that by definition, $\{ \Phi_{\K}(\{\r_i\}) \}$ in different momentum sectors are mutually orthogonal.    

Topological order implies the existence of an $m$-th dimensional (quasi)-degenerate ground state manifold of $H$ where $m$ is the topological degeneracy. In the presence of translational symmetry, the states within this manifold are labeled by COM momentum $\K^{\rm top}_{i}$ with $i = 1, \cdots, m$. Generally, the specific COM momentum values depend on the geometry of the finite torus \cite{regnault2011fractional,pu2017composite}. Here, we consider geometries where each of the  ground states has a different value of $\K^{\rm top}_{i}$ and comment later on other cases where some (or all) of the ground states lie in the same momentum sector. 

The core of our protocol is the following:  If $\Psi(\{\r_i\})$ is an accurate approximation to the true 
ground state of $H$, (1) it should have non-zero weights $\{c_{\K}\}$ \textit{only} for values of $\K \in \{\K_i^{\rm top}\}$ corresponding to the momentum sectors of the 
degenerate ground states. (2) The variational energies $ E_{\K} = \langle \Phi_{\K}|H|\Phi_{\K} \rangle$ with $\K \in \{{\K_i^{\rm top}}\}$ should be nearly-degenerate if the energy of $\Psi(\{\r_i\})$ is close to the true ground state energy.  

\begin{figure}
    \centering
    \includegraphics[width=0.85\linewidth]{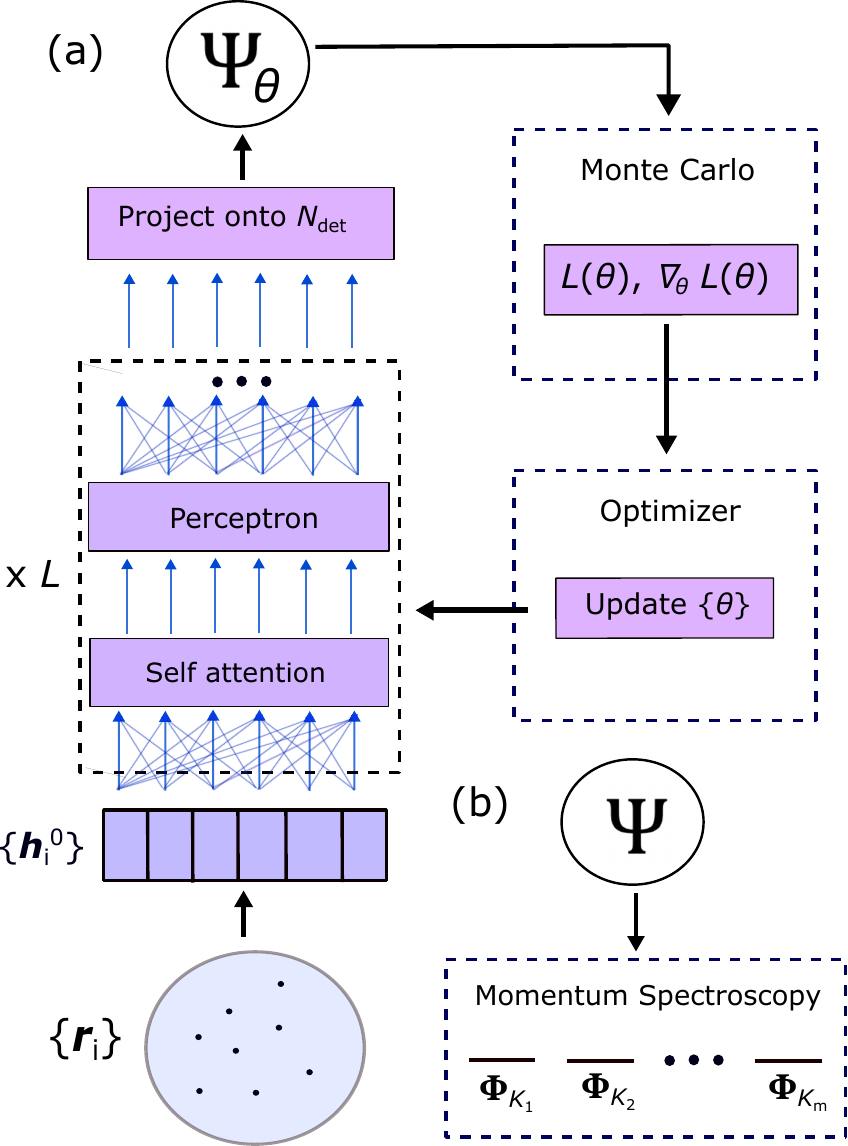}
    \caption{(a) Schematic of the self-attention–based neural network architecture. Particle positions ${\r_i}$ are mapped to high-dimensional vectors ${\h_i^0}$, which are processed by $L$ layers of self-attention and perceptron transformations (Appendix \ref{AppnA}). The final output is projected onto $N_{\rm det}$ determinants to form the variational ansatz \eqref{eq:ansatz}. The parameters ${\theta}$ are optimized via Monte Carlo sampling of the variational energy
$L(\theta) = \langle \Psi_{\theta}|H|\Psi_{\theta}\rangle / \langle \Psi_{\theta}|\Psi_{\theta} \rangle$. (b) The optimized wavefunction is decomposed to different momentum sectors at the end of optimization. }
    \label{Fig:NNschematic}
\end{figure}

Therefore given a variational wavefunction $\Psi(\{\r_i\})$, the existence of topological order can be inferred as described above (Fig. \ref{Fig:NNschematic}(b)). To demonstrate the validity and usefulness of our method, we apply it to the ground states of fractional Chern insulators. 

\section{Minimal Model} 

First, we introduce a minimal continuum model which describes spinless fermions with parabolic dispersion in a periodic magnetic field. The Hamiltonian takes the form,
\begin{equation}
\label{eq:minimalmodel}
    H = \sum_{i} \dfrac{(-i \hbar \boldsymbol{\nabla}_i + e \A(\r_i))^2}{2 m} + \sum_{i < j} \dfrac{e^2}{4 \pi \epsilon |\r_i - \r_j|}
\end{equation}

where $\A(\r)$ is the vector potential of a periodically varying magnetic field $B(\r) = \curl  \A(\r)$. The second terms  describes Coulomb interactions between the particles controlled by the dielectric constant $\epsilon$.

Our model is motivated by the continuum Hamiltonian of twisted TMDs \cite{wu2019topological}, which feature a spatially varying Zeeman field acting on the layer pseudospin degree of freedom that arises from the underlying moiré intralayer potential and interlayer tunneling. 
In the adiabatic limit of sufficiently strong Zeeman coupling, the layer pseudospin locally aligns with the Zeeman field, and the problem maps onto a Hamiltonian of holes in a periodic magnetic field and periodic scalar potential \cite{paul2023giant, morales2024magic,shi2024adiabatic}. Crucially, in this mapping, the net flux per moiré unit cell is determined by the topological winding number $\mathcal{N}$ of the layer-pseudospin texture, $  \frac{1}{\Phi_0}  \int_{\rm UC} d^2 \r B(\r) = \mathcal{N}$.

Previous studies \cite{morales2024magic,shi2024adiabatic,li2025variational,reddy2024non,reddy2024anti} have focused on the regime where the pseudospin texture forms a skyrmion lattice in real space, corresponding to winding number $\mathcal{N} = 1$. In this regime, 
twisted TMDs featuring an emergent magnetic field closely resemble Landau levels under a real magnetic field, with the field strength corresponding to one flux quantum per moir\'e unit cell area. This resemblance naturally explains the observed sequence of fractional Chern insulators that are the analogs of Jain sequence fractional quantum Hall states.      

Here, we instead focus on a different regime $\mathcal{N} = 0$, corresponding to zero average flux per unit cell. This is realized when the displacement field is strong enough that the holes are predominantly localized in a single layer. Importantly, as we shall see below, vanishing flux per unit cell does not necessarily imply a topologically trivial ground state. 

For concreteness, we choose a $C_6$ symmetric magnetic field, a sum of the lowest harmonics, \begin{equation}
    B(\r) = 2 B_0 \sum_{i = 1,2,3} \cos(\b_i \cdot \r) 
\end{equation}
$\b_1=b_0 (1/2,\sqrt{3}/2)$, $\b_2=-b_0(1,0)$ and $\b_3=-\b_1-\b_2$ with $b_0= 4 \pi/\sqrt{3} a_0 $ with $a_0$ the underlying lattice constant of the magnetic field unit cell. In what follows, we define the dimensionless parameter $r_s = 1/\sqrt{\pi n} a_B$ where $n$ is the density and $a_B$ is the Bohr radius, $a_B = 4 \pi \epsilon \hbar^2/m^2$. $r_s$ provides a measure of the ratio  of the interaction strength to the bare kinetic energy. In addition, we define another dimensionless parameter $\lambda = \frac{2 \pi}{ \Phi_0} \frac{B_0}{b_0^2}$ with $\Phi_0 = h /e$ is the magnetic flux quantum. $\lambda$ measures the fluctuations of the periodic magnetic field. 

\begin{figure}
    \centering
    \includegraphics[width=\linewidth]{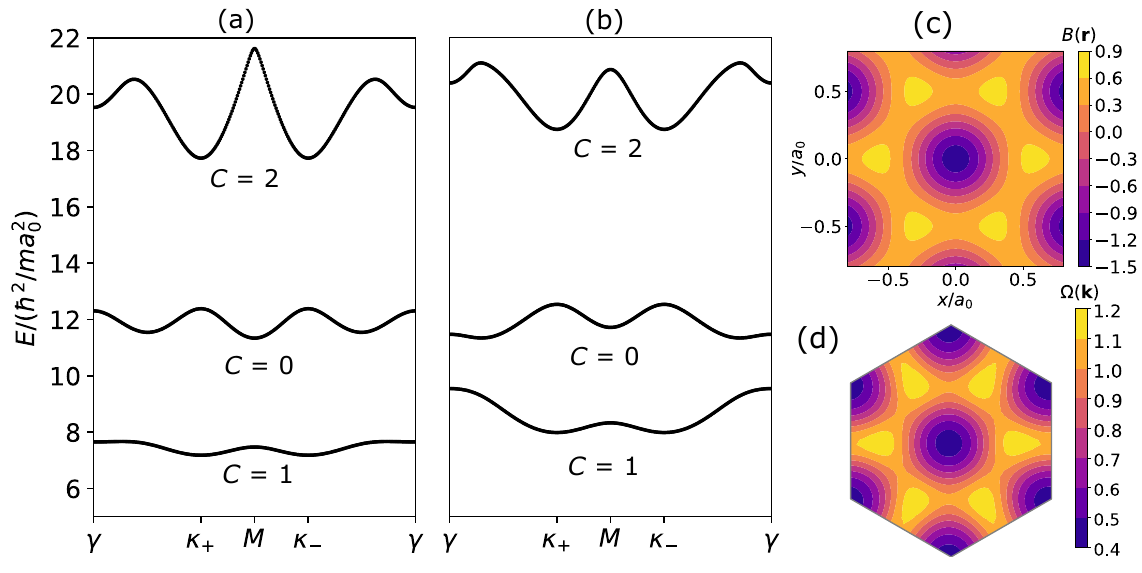}
    \caption{Band structure of the model defined in equation \eqref{eq:minimalmodel} obtained for (a) $\lambda = -0.23$ and (b) $\lambda = -0.26$. (c) The periodic magnetic field in real space for $\lambda = -0.23$. (d) Berry curvature of the lowest band for $\lambda = -0.23$. }
    \label{Fig1}
\end{figure}

In Fig. \ref{Fig1}(a), we show the non-interacting band structure when the magnetic field is highly non-uniform. We find the lowest band to be very flat and carries a Chern number $C = 1$ for a wide range of $\lambda$ values. Interestingly, the Berry curvature is  highly fluctuating across the Brillouin zone (Fig. \ref{Fig1}(d)). The flatness of the lowest band implies that interaction effects will be very important and could lead to topologically ordered phases, a possibility which we investigate next.




\section{Neural Network Variational Monte Carlo}

We study the Hamiltonian \eqref{eq:minimalmodel} with variational Monte Carlo (VMC). Our variational ansatz for the many-body ground state is given by
\begin{equation}
\label{eq:ansatz}
    \Psi_{\{\theta\}}(\{\r_i\})  = \sum_{n}^{N_{\rm det}} {\rm det}[\phi_j^n(\r_i; \{\r_{\neq i }\})]
\end{equation}
where $\phi_j^n(\r_i; \{\r_{\neq i }\})$ describes a \textit{generalized} many-body orbital for the $i$-th particle that depends on the coordinates of all other particles in the system, $ \{\r_{\neq i }\} \equiv \{\r_{k \neq i}\}$. $\{\theta\}$ denote the variational parameters of the ansatz. 

In this paper, we parametrize the many-body orbitals $\phi_j^n(\r_i; \{\r_{\neq i }\})$ by a deep neural network \cite{pfau2020ab}. In particular, we utilize the Psiformer architecture \cite{von2022self,geier2025self} which is based on the self-attention mechanism.  Such an architecture has been quite successful in describing highly non-trivial quantum phases of matter such as fractional quantum Hall systems \cite{teng2025solving,nazaryan2025artificial} and chiral superconductors \cite{li2025attention}. The details of the architecture (schematically shown in Fig. \ref{Fig:NNschematic} (a)) can be found in the Appendix \ref{AppnA}.


 Our VMC calculations are performed on a real space torus  spanned by two vectors $\L_1$ and $\L_2$. To respect the periodic boundary conditions of the torus, the neural network takes  \textit{periodized} electron coordinates [$\sin(\G_a \cdot \r_i)$ and $\cos(\G_a \cdot \r_i)$] as inputs where $\G_{a = 1,2}$ are the reciprocal lattice vectors of the torus, i.e,  $\G_a \cdot \L_b = 2\pi \delta_{ab} $.


The parameters $\{\theta\}$ of the neural network which describe the  ansatz \eqref{eq:ansatz} are optimized (trained) to minimize the loss function $L(\theta)$ corresponding to the total energy $L(\theta) =  \langle \Psi_{\theta}|H|\Psi_{\theta}\rangle / \langle \Psi_{\theta}|\Psi_{\theta} \rangle $(Appendix \ref{AppnB}), yielding a variational approximation to the many-body ground state.

Importantly, no prior knowledge of the system such as the band structure or band topology (Fig.~\ref{Fig1}) is built into the NN ansatz before optimization. Additionally, no pretraining is performed. All the parameters are instead randomly initialized, leading to very high initial  energy (Appendix \ref{AppnD}). Remarkably, as we show next, the NN systematically converges to a topologically ordered ground state without any input bias.


\begin{figure}[t!]
    \centering
    \includegraphics[width=1.0\linewidth]{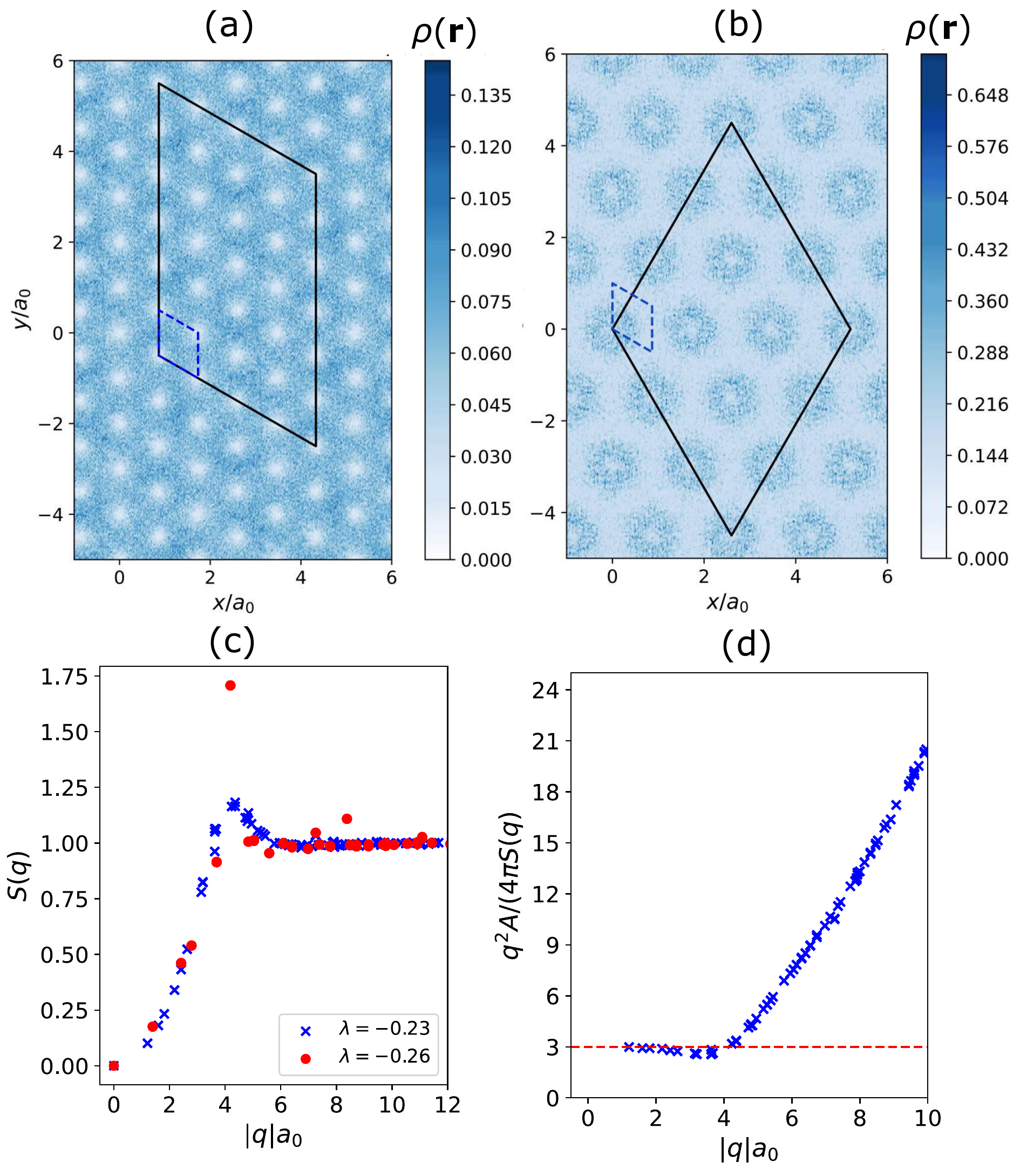}
    \caption{Ground state charge density $\rho(\r)$ at $\nu = 1/3$ for (a) the FCI phase for $\lambda = -0.23$ and (b) the CDW phase for $\lambda = -0.26$. The solid black parallelogram denotes the supercell while the dashed blue parallelogram denotes the unit cell of the magnetic field. In (a), we use a supercell with 24 unit cells (8 particles) and in (b), we use a supercell with 27 unit cells (9 particles) (c) Structure factor $S(\q)$ as a function of $|\q|$ computed in the FCI phase ($\lambda = -0.23$) and the CDW phase ($\lambda = -0.26$). (d) $|\q|^2 A / 4 \pi S(\q)$ in the FCI phase ($\lambda = -0.23$) approaches $3$ for small $|\q|$, almost saturating the topological bound \cite{onishi2024topological}. All calculations are done for $r_s \approx 3.43$. }
    \label{fig:chargedensity}
\end{figure}

\section{Results} 

First, we focus on filling factor $\nu = 1/3$ in the presence of strong magnetic field modulations $\lambda = -0.23$ which give rise to the flat Chern $C = 1$ band (Fig. \ref{Fig1}(a)).  We find the optimized NN wavefunction at this filling factor to exhibit density features consistent with a translationally-symmetric liquid-like state, as evident in the real space density $\rho(\r)$ in Fig. \ref{fig:chargedensity}(a). In addition, we compute the static structure factor $S(\q) = \frac{1}{N} \langle\rho_{\q} \rho_{-\q} \rangle$ with $\rho_{\q} = \sum_i e^{i \q \cdot \r_i}$  the density operator and $N$ the number of particles.  As shown in Fig. \ref{fig:chargedensity}(c), $S(\q)$ does not show any Bragg peaks, further confirming the liquid nature of the state.

On the other hand, as the strength of magnetic field modulations becomes stronger ($\lambda = -0.26$), the lowest Chern $C = 1$ band remains topological (Fig. \ref{Fig1}(b)) but we observe a transition to a charge density wave phase as shown in Fig. \ref{fig:chargedensity}(b) where the unit cell is tripled corresponding to a $\sqrt{3} \times \sqrt{3}$ periodicity. This periodicity is also manifested in $S(\q)$ (Fig. \ref{fig:chargedensity} (c)) which exhibits pronounced peaks at the corners of the Brillouin zone, $\q = \kappa_{+}, \kappa_{-}$ with $|\q| = \frac{4 \pi}{3 a_0} $.

In both phases, NN-VMC achieves lower energy than ED projected onto the lowest band (Appendix \ref{AppnD}). Because NN-VMC works directly in real space without band projection, it naturally includes all bands, whereas multiband ED is beyond computational reach for the system sizes we study here.

\begin{figure}
    \centering
    \includegraphics[width=\linewidth]{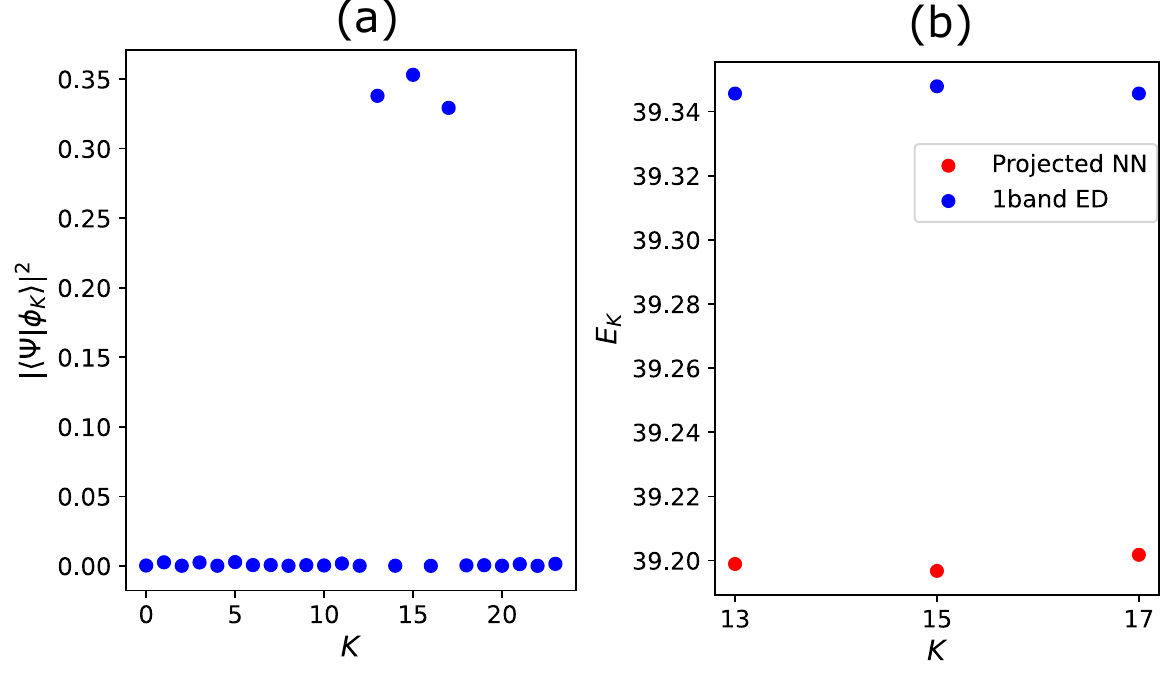}
    \caption{(a) The overlap $|\langle \Psi(\{\r_i\})| \Phi_{\K}(\{\r_i\})\rangle|^2$ of the optimized wavefunction $\Psi(\{\r_i\})$ with its projection $\Phi_{\K}(\{\r_i\})$ onto different COM momentum sectors $K$. (b) Variational energy $E_{\K} = \langle \Phi_{\K}|H|\Phi_{\K} \rangle / \langle \Phi_{\K}|\Phi_{\K} \rangle$  (in units of $\hbar^2/m a_0^2$) of the wavefunction in the three different momentum sectors with non-zero weight (shown in (a)).}
    \label{fig:NNenergy}
\end{figure}

To gain more insights about possible topological order in the system for $\lambda = -0.23$, we investigate the small $\q$ behavior of the full structure factor $S(\q)$ which takes the form, $S(\q) \approx K |\q|^2 + O(|\q|^4)$ for gapped systems. It has been shown \cite{onishi2024topological,zaklama2025structure} that the quantum weight $K$ is bounded from below: $K \geq \frac{A |C|}{4 \pi}$ where $C$ is the many-body Chern number. In Fig. \ref{fig:chargedensity}(d), we find $ |\q|^2 A / 4 \pi S(\q)$ to approach 3 for small $|\q|$ implying that the bound is not violated and the ground state is consistent with being a topological phase. In addition, for a gapped phase, this quantity is directly related to the collective mode dispersion obtained from the single mode approximation $\Delta(\q) = \frac{\hbar^2 |\q|^2}{2 m S({\q})}$ \cite{feynman1954atomic} . As $S(\q)$ is unprojected, the gapped liquid's collective mode comprises both intra- and inter-band excitations. 

Having ruled out the possibility of crystalline order for $\lambda = -0.23$ based on the ground state charge density and structure factor, we apply the post-processing protocol we described above to diagnose possible topological order. The NN wavefunction $\Psi(\{\r_i\})$ is projected onto all crystalline momentum sectors to obtain  $\Phi_{\K}(\{\r_i\})$ (Eq. \eqref{eq:projection}). We then compute the overlap $|\langle \Psi(\{\r_i\})| \Phi_{\K}(\{\r_i\})\rangle|^2$ to extract the weights of the NN wavefunction in the different momentum sectors. As shown in Fig. \ref{fig:NNenergy}(a), the wavefunction has non-zero weight \textit{only} in three sectors. Remarkably, these momentum sectors correspond to the FCI ground state momenta for this finite system which can be deduced from counting rules based on generalized Pauli principles or thin-torus limits \cite{regnault2011fractional,bergholtz2008quantum,bernevig2008model,bernevig2012emergent}.

Furthermore, we find the variational energy $E_{\K} = \langle \Phi_{\K}|H|\Phi_{\K} \rangle / \langle \Phi_{\K}|\Phi_{\K} \rangle$ for each of the three momentum sectors to be quasi-degenerate (Fig. \ref{fig:NNenergy} (b)) reflecting the underlying topological degeneracy. In addition, these energies are significantly lower than the corresponding energies obtained from band projected ED. 

\section{Discussion} 

In this paper, we have demonstrated the ability of neural network quantum states based on the self-attention mechanism to capture topologically ordered fractional Chern insulators only through energy minimization without any input bias. Remarkably, our method achieves significantly lower energies than band projected exact diagonalization while being able to capture arbitrary band mixing effects. Using the same ansatz, the neural network is able to capture competing charge density waves.

We have introduced a diagnostic to reveal ground-state topological degeneracy directly from NN real-space wavefunction. Crucially, our approach differs from Ref.~\cite{li2025deep} as we do not enforce translation symmetry during optimization. Instead, we project onto momentum sectors only once—at the end—using a single optimized wavefunction. This avoids the substantial computational cost of separate optimizations in each momentum sector which requires translating the ansatz $N_s$ times \cite{zhang2025neural}.

Our construction can unambiguously  detect topological degeneracy if the ground states lie in different momentum sectors. However, if some (or all) of the ground states share the same momentum quantum numbers, our approach can still be utilized if these states carry distinct quantum numbers under different symmetries such as rotation  and inversion \cite{kobayashi2025crystalline,fu2007topological}. Additionally, this approach is not unique to topologically ordered states but can be used to detect generic low-lying excitations above the ground states \cite{choo2018symmetries}.

We have studied a model of spinless fermions in a periodic magnetic field with zero net flux per unit cell. In addition to twisted semiconductors, such a model can also be realized in multi-valley electron systems subject to periodic strain \cite{tang2014strain,tahir2020emergent,gao2023untwisting} and   
ultracold-atom experiments that implement synthetic gauge fields \cite{aidelsburger2013realization}.


Having established that self-attention neural-network quantum states can describe abelian topological order, a natural future direction is to use NN-VMC to investigate non-Abelian topological phases \cite{abouelkomsan2025non,reddy2024non} and gapless states such as the composite Fermi liquid (CFL).

\section{Acknowledgments}
It is a pleasure to acknowledge useful discussions with Aidan Reddy, Daniele Guerci and Pierre-Antoine Graham. This material is based upon work supported by the Air Force Office of Scientific Research under award number FA2386-24-1-4043.
A. A. was supported by the Knut and Alice Wallenberg Foundation (KAW 2022.0348). L. F. was supported in part by a Simons Investigator Award from the Simons Foundation. This work made use of resources provided by subMIT at MIT Physics.

\setcounter{figure}{0}
\setcounter{section}{0}
\setcounter{equation}{0}
\setcounter{NAT@ctr}{0}

\makeatother

\appendix 

\section{Neural Network Architecture}

\label{AppnA}
In this section, we describe the neural network architecture used throughout the paper. We will follow closely Ref. \cite{geier2025self}. We use an architecture that is schematically shown in Fig. 2 in the main text and has both self-attention and multilayer perceptron components. As we are working with a torus supercell, the wavefunction satisfies periodic boundary conditions $\Psi(\r_1, \cdots, \r_i + \L_{a}, \cdots, \r_N) = \Psi(\r_1, \cdots, \r_N) $ for all particles $\r_i$ where $\L_{a}$ with $ a= 1,2$ denote the two primitive supercell vectors. First a feature vector $f_i \equiv f(\r_i)$ is defined for the $i$-th particle as \begin{equation}
    \f_i \equiv f(\r_i) = \begin{pmatrix}
        \cos(\G_1 \cdot \r_i) \\
        \cos(\G_2 \cdot \r_i) \\
        \sin(\G_1 \cdot \r_i) \\
        \sin(\G_2 \cdot \r_i)
    \end{pmatrix}
\end{equation}
This feature vector represents a periodized version of the particle coordinates then each feature vector is mapped to an initial $\h_i^{0}$ vector which lives in a higher dimensional space through a transformation, \begin{equation}
    \h_i^{0} = W^{0} \f_i 
\end{equation}
with $W^0 \in \mathbb{R}^{d_L} \times \mathbb{R}^{2 d_{\rm dim}}$ with $d_L$ is the dimension of the higher dimensional space where the particle coordinates are embedded which corresponds to the width of the internal layers of the neural network. $d_{\rm dim}$ is the spatial dimension ($d_{\rm dim} = 2$ in our case). Importantly, $W^0$ is the same for each particle coordinate. Having mapped the coordinates of all particles to vectors $\{\h_i^0\}$, these vectors undergo two different types of transformations, multi-head self attention followed by multilayer perceptron (MLP) which we explain next. 
\subsection{Multilayer Perceptron (MLP)}
MLPs are standard feed-forward neural networks that implement the following transformation on a \textit{generic} vector $\g_1 \in \mathbb{R}^{d_L}$ to yield another vector $\g_2$, \begin{equation}
\label{eq:MLP}
    \g_2 = \g_1 + F(W \g_1 + \b)
\end{equation}
with $W$ is a linear transformation $W \in \mathbb{R}^{d_L} \times \mathbb{R}^{d_L}$ and $\b \in \mathbb{R}^{d_L}$ is a bias vector. $F$ here represents a non-linear activation function which we choose to be the GELU function. 
\subsection{Self-attention}
In the NN architecture (Fig. 2 in the main text), MLPs act individually on each particle without mixing different particle streams which cannot describe correlations between particles. In order to capture such correlations, we utilize the self-attention mechanism \cite{vaswani2017attention} which form the basis of transformers used in large language models. Self-attention mechanism learns how each particle is influenced by the remaining particles. Self attention acts of all set of particle streams in the $l$-layer $\{\h_i^{l}\}$ to generate intermediate streams $\{\g_i^{l}\}$ that are then passed to the MLP \eqref{eq:MLP}. Schematically we have the following, 
\begin{equation}
    \cdots \rightarrow \{\h_i^{l}\} \xrightarrow{\rm SELF-ATTN}  \{\g_i^{l}\} \xrightarrow{\rm PERCEPTRON} \{\h_i^{l+1}\} \rightarrow \cdots 
\end{equation}

In the self-attention mechanism, three features for each element in $\{h_i^{l}\}$ are defined which are called \textit{keys}, \textit{queries} and \textit{values} and are defined as,
\begin{equation}
    \k_i^{l h} = W_{k}^{lh} \h_i^{l}, \> \> W_{k}^{lh} \in \mathbb{R}^{d_{\rm attn}} \times \mathbb{R}^{d_L}
\end{equation}
\begin{equation}
    \q_i^{l h} = W_{q}^{lh} \h_i^{l}, \> \> W_{q}^{lh} \in \mathbb{R}^{d_{\rm attn}} \times \mathbb{R}^{d_L}
\end{equation}
\begin{equation}
    \v_i^{l h} = W_{v}^{lh} \h_i^{l}, \> \> W_{v}^{lh} \in \mathbb{R}^{d_{\rm attnvals}} \times \mathbb{R}^{d_L}
\end{equation}
The transformations $\{W_{k}^{lh},W_{q}^{lh}, W_{v}^{lh} \} $ are the same for each particle $i$. We use multi-head attention where more than one self-attention operation is applied, indexed by $h$ in $W_{k}^{lh}$ , $W_{q}^{lh}$ and $ W_{v}^{lh}$ which is not to be confused with the particle stream vectors $\h_i^{l}$. The dimensions $d_{\rm attn}$ and $d_{\rm attnvals}$ are different from $d_{L}$ and are generally much smaller.  The keys and queries vectors have the same vector space dimension $d_{\rm attn}$ to allow for comparison between two different particle streams $i,j$ through the dot product $\k_i^{lh} \cdot \q_j^{lh}$. The values $\v_i^{lh}$ is a measure of the influence on particle $i$ from the remaining particles $j \neq i$. The value of self-attention for each particle $i$ is computed as 
\begin{equation}
    \begin{aligned}
\label{eq:selfattn}
    {\rm SELFATTN}_i(\{\h_j^l\}; W_{k}^{lh},W_{q}^{lh}, W_{v}^{lh}) \\ = \frac{1}{\mathcal{N}} \sum_{j = 1}^N {\rm exp}(\k_i^{lh} \cdot \q_j^{lh}) \v_j^{lh} 
\end{aligned}
\end{equation}
with a normalization factor, \begin{equation}
    \mathcal{N} = \sqrt{d_{\rm attnvals}} \sum_{j = 1}^N {\rm exp}(\k_i^{lh} \cdot \q_j^{lh})
\end{equation}
The physical meaning of \eqref{eq:selfattn} is transparent. The self-attention value of particle $i$ returns the value $\v_j^{l h}$ of particle stream $j$ weighted by the exponential factor which measures the correlations between $i,j$ such that the most relevant particles $j$ for each particle $i$ are identified. The intermediate streams are given by 
\begin{equation}
\begin{aligned}
    & g_i^{l+1} = \h_i^{l} + W_o^{l} \\ & \times \> {\rm concat}_h \> [{\rm SELFATTN}_i(\{\h_j^l\}; W_{k}^{lh},W_{q}^{lh}, W_{v}^{lh})]
    \end{aligned}
\end{equation}
All values of self-attention from the multi-head attention are concatenated then transformed through transformation $W_0^{l} \in \mathbb{R}^{d_L} \times \mathbb{R}^{N_{\rm heads} \times {\rm attnvals}}$. The intermediate streams $\{\g_i^{l}\}$ are passed through MLP giving rise to,
\begin{equation}
    \h_i^{l+1} = \g_i^{l} + F(W^{l} \g_i^{l} + \b^{l+1})
\end{equation}
Here $W^l$ and $\b^{l}$ are the same for all particle streams which are labeled by $i$. 
\subsection{Projection onto $N_{\rm det}$}
The final output of the neural network $\{\h_i^L\}$ is projected onto $n = 1, \cdots, N_{\det} $ distinct sets of generalized many-body orbitals. \begin{equation}
    \phi_j^n(\r_i; \{\r_{\neq i }\}) = \w_{2 j}^n \cdot \h_i^{L} + i \w_{2 j+1}^n \cdot \h_i^{L}
\end{equation}
where $\w_{2j}^n$ and $\w_{2j+1}^n$ are projection matrices that construct the real and the imaginary part of the generalized orbitals. The full wavefunction ansatz reads,
\begin{equation}
\label{eq:NNansatz_SM}
        \Psi_{\{\theta\}}(\{\r_i\})  = \sum_{n}^{N_{\rm det}} {\rm det}_{ij} \phi_j^n(\r_i; \{\r_{\neq i }\})
\end{equation}
where $\{\theta\}$ here represent all parameters of the ansatz which include the initial embedding matrix $W^0$, the self attention matrices $\{W_{k}^{lh},W_{q}^{lh}, W_{v}^{lh}, W_o^l \}$, the MLP transformations $\{W^l, \b^l\}$ and the final projection matrices $\{\w_{2j}^n, \w_{2j+1}^n\}$.
It is important to note that in order for the ansatz \eqref{eq:NNansatz_SM} to describe an fermionic anti-symmetric wavefunction, the generalized orbitals $\phi_j^n(\r_i; \{\r_{\neq i }\})$ have to be permutation equivariant in the coordinates $\{\r_{\neq i }\}$,
\begin{equation}
    \phi_j^n(\r_i; \{\r_k, \cdots, \r_l,\cdots\}) =  \phi_j^n(\r_i; \{\r_l, \cdots, \r_k,\cdots\})
\end{equation}
for $k,l \neq i$. This property is inherited from the form of the self-attention operations \eqref{eq:selfattn}. In Table. \ref{Tab:Hyperparams}, we list the hyperparameters of the self-attention neural network we use.

\begin{table}
    \centering
    \renewcommand{\arraystretch}{1.3}
    \begin{tabular}{ll l}
        \toprule
        \textbf{Parameter} & & \textbf{Value} \\
        \midrule
        \textbf{Architecture} & & \\ 
        Network layers & & $L = 4$ \\
        Attention heads per layer & & $N_{\rm heads} = 6$ \\
        Attention dimension & & $d_{\rm attn} = d_{\rm attnvals} = 64$ \\
        Perceptron dimension & & $d_L = 256$ \\
        $\#$ perceptrons per layer & & $2$ \\
         Determinants & & $N_{\rm det} = 4$\\
         Layer norm & & True \\
        \midrule
        \textbf{Training} & & \\ 
        Training iterations & & $10e6 - 15e6$ \\
        Learning rate at time $t$  & & $\eta = \eta_0(1 + \frac{t}{t_0})^{-1}$ \\
        Initial learning rate &  & $\eta_0 = 0.001$ \\
        Learning rate delay &  & $t_0 = 1e6$ \\
        Local energy clipping &  & $\rho = 5.0$ \\
        \midrule
        \textbf{MCMC} & & \\
        Batch size & & $1024$ \\
        Initial move width & & $0.2$ \\
        \midrule
        \textbf{KFAC} & & \\ 
        Norm constraint & & $1 \times 10^{-3}$ \\
        Damping & & $1 \times 10^{-3}$ \\
        $L_2$ regularization & & 0.0 \\
        Momentum & & 0.0 \\
        \bottomrule
    \end{tabular}
    \caption{Table of default hyperparameters used in our numerical calculations with the self-attention neural network.}
    \label{Tab:Hyperparams}
\end{table}

 \section{Optimization}
 \label{AppnB}
 The variational parameters $\{\theta\}$ of the wavefunction ansatz \eqref{eq:NNansatz_SM} are optimized (trained in the language of AI) to minimize the energy. In VMC, the total energy is computed as,
 \begin{equation}
     \begin{aligned}
\label{eq:totalenergy_SM}
    E_{\{\theta\}} = \dfrac{\int \{d \r_i\} \Psi^{*}_{\{\theta\}}(\{\r_i\}) H \Psi_{\{\theta\}}(\{\r_i\})}{\int \{d \r_i\} \Psi^{*}_{\{\theta\}}(\{\r_i\}) \Psi_{\{\theta\}}(\{\r_i\})} \\ = \mathbb{E}_{\{\r_i\} \sim p_{\theta}}(E^L_{\{\theta\}}) \approx \dfrac{1}{M} \sum_{M} E^L_{\{\theta\}}(\{\r_i\}) 
\end{aligned}
 \end{equation}

The total energy is given as the expectation value of the \textit{local} energy $E^L_{\{\theta\}}(\{\r_i\}) = \Psi^{-1}_{\{\theta\}}(\{\r_i\}) H \Psi_{\{\theta\}}(\{\r_i\}) $ over probability distribution defined by the ansatz $p_{\theta} =  \frac{|\Psi_{\{\theta\}}(\{\r_i\})|^2}{\int \{d \r_i\} \Psi^{*}_{\{\theta\}}(\{\r_i\}) \Psi_{\{\theta\}}(\{\r_i\})}$. In the last line of \eqref{eq:totalenergy_SM}, this expectation value is approximated as the average of the local energy computed over a number of (statistically independent) samples $M$ of configurations of particles $\{\r_i\}$ that are drawn according to $p_{\theta}$ using Markov Chain Monte Carlo (MCMC).

For optimization purposes, the gradient of $E_{\{\theta\}}$ with respect to $\{\theta\}$ needs to be computed and it is given by \cite{becca2017quantum}, 
\begin{equation}
    g_{a} \equiv \partial_{\theta_a} E_{\{\theta\}} =  2 \mathbb{E}_{\{\r_i\} \sim p_{\theta}} \big[(E^L_{\{\theta\}} - E_{\{\theta\}})  \partial_{\theta_a} {\rm log} \> \Psi_{\{\theta\}} \big]
\end{equation}
where $a$ denotes the $a$-th component of the vector of the parameters $\boldsymbol{\theta} \equiv \{\theta\}$. 

The parameters $\boldsymbol{\theta}$ are generally updated in a gradient descent manner, \begin{equation}
\label{eq:updaterule}
    \theta^{t+1}_{a} = \theta_{a}^{t} - \eta_t \> \alpha^{-1}_{a b} \> g_b
\end{equation}
where $t$ denotes the $t$-th optimization iteration. $\eta_t$ is known as the learning rate which sets the degree of descent along the gradient $g_{a}$ and $\alpha_{ab}$ is a preconditioning matrix. The product $\> \alpha_{a b} \> g_b$ is understood to be evaluated at $\theta^{t}_{a}$. 
The choice of the preconditioning matrix $\alpha_{ab}$ is crucial and corresponds to various gradient descent algorithms.  For example, if $\alpha_{ab} = \delta_{ab}$ then the parameter update corresponds to standard first order gradient descent which does not use any second order information of the parameter manifold (such as the curvature) while if $\alpha_{ab} = H_{ab}$ where $H_{ab} = \partial_{\theta_a} \partial_{\theta_b}  E_{\{\theta\}} $ is the full Hessian matrix, the parameter update corresponds to second order Newton's method which is guaranteed to converge to a minimum. 



Here, we use the KFAC optimizer \cite{martens2015optimizing}, a quasi-second order method that typically performs better than commonly first order methods in NN optimization such as ADAM \cite{adam2014method}.

\section{Model}

\begin{figure}
    \centering
    \includegraphics[width=1.0\linewidth]{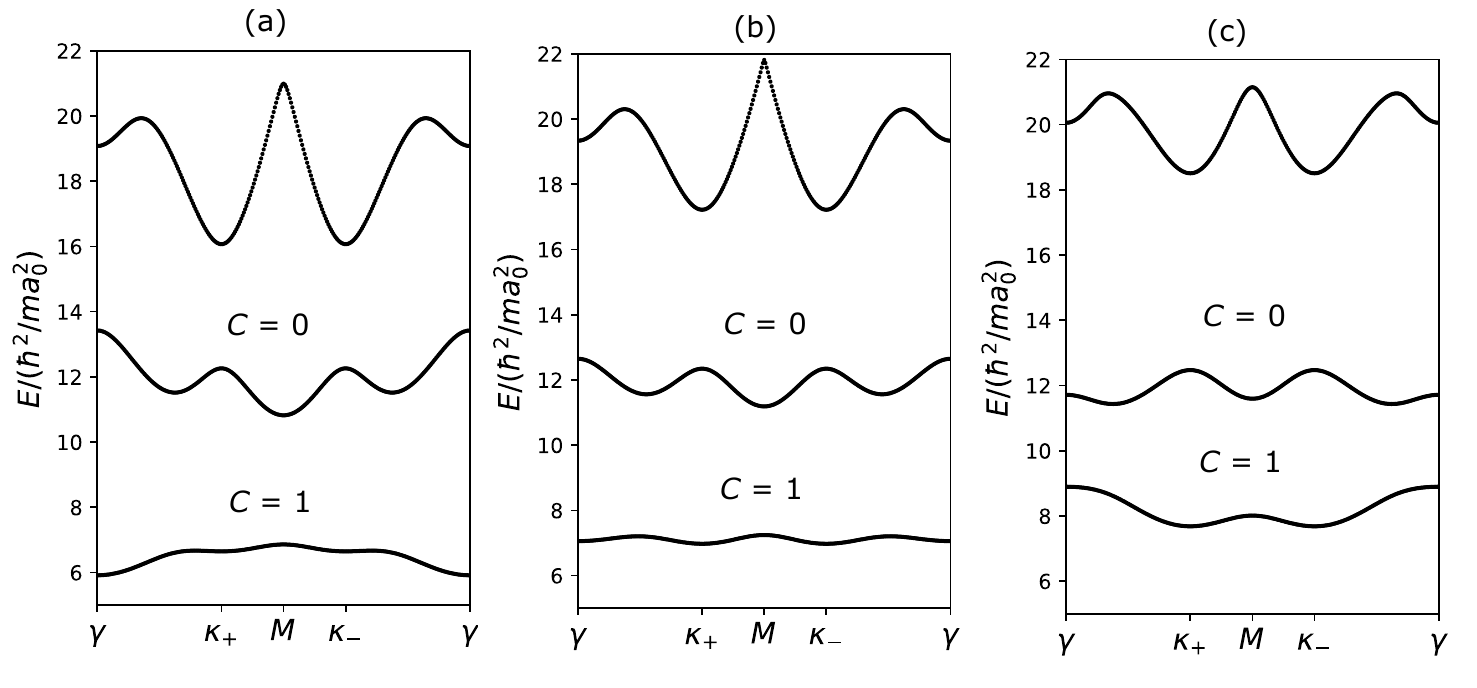}
    \caption{Band structure of the model \eqref{eq:minimalmodel_SM} for (a) $\lambda = -0.2$, (b) $\lambda = -0.22$ , (c) $\lambda = -0.25$}
    \label{fig:bandstructure_combined}
\end{figure}

As detailed in the main text, we study a model of spinless fermions in a periodic magnetic field with net average flux of zero in the unit cell \cite{tahir2020emergent}. This model bears resemblance to the Haldane model, formulated in the model. However the band structure and topology are quite different. We copy the non-interacting Hamiltonian below, 
\begin{equation}
\label{eq:minimalmodel_SM}
    H =  \dfrac{(-i \hbar \boldsymbol{\nabla} + e \A(\r))^2}{2 m} 
\end{equation}
We choose a $C_6$ symmetric magnetic field,
\begin{equation}
 B(\r) = 2 B_0 \sum_{i = 1,2,3} \cos(\b_i \cdot \r) 
\end{equation}
$\b_1=b_0 (1/2,\sqrt{3}/2)$, $\b_2=-b_0(1,0)$ and $\b_3=-\b_1-\b_2$ with $b_0= 4 \pi/\sqrt{3} a_0 $ with $a_0$ the underlying lattice constant of the magnetic field unit cell. 
We define the dimensionless parameter $\lambda = \frac{2 \pi}{ \Phi_0} \frac{B_0}{b_0^2}$ with $\Phi_0 = h/e $ is the magnetic flux quantum. $\lambda$ measures the strength of the magnetic field fluctuations. In Fig. \ref{fig:bandstructure_combined}, we plot the evolution of the band structure as a function of $\lambda$. We find a flat Chern band with $C = 1$ which becomes more flat as $\lambda$ increases. However, as $\lambda$ increases, we observe a transition from FCI at $\nu = 1/3$ to a CDW as discussed in the main text. We leave a comprehensive investigation of this model to future work.

\section{Comparison with Exact Diagonalization}
\label{AppnD}
\begin{figure}
    \centering
    \includegraphics[width=0.9\linewidth]{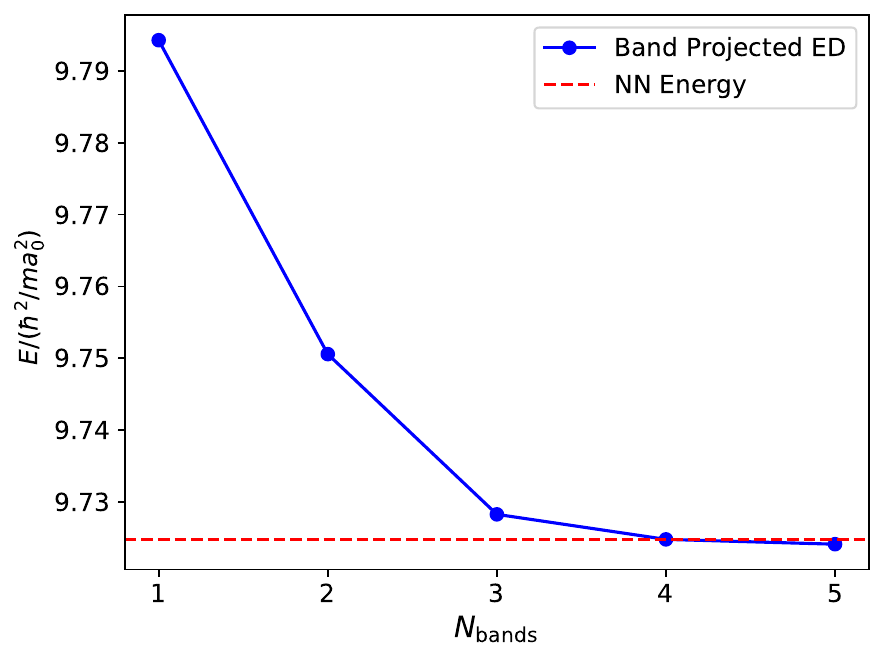}
    \caption{Comparison of the variational energy of the NN wavefunction to the lowest energy obtained from exact diagonalization projected onto $N_b$ bands. Calculations were done on a supercell with $3 \times 3$ unit cells at $\nu= 1/3$ for $r_s \approx 3.43$ and $\lambda = -0.23$ }
    \label{fig:NNvsED}
\end{figure}

Here, we compare the energies obtained with NN-VMC to the energies obtained from band projected ED. As shown in Fig. \ref{fig:NNvsED}, the NN energies are significanly lower than ED projected onto $N_b = 1,2,3$ bands for a small system with three particles in 9 unit cells. Increasing the number of bands $N_b$, the ED energies approache the true energy value of the systems. However, the NN energy is still very close showing that it is an  excellent approximation for the ground state of the system.

In Table. \ref{tab:NNenergiesvsED}, we list the energy of the NN wavefunction (unprojected onto any momentum sector) compared with the energy obtained from 1 band ED. Again, we find the NN to exhibit lower energies. For these system sizes, multiband ED is beyond computational reach. 

\begin{table}
  \centering
  \begin{tabular}{|l|c|c|}
    \hline
     Parameters & 1 band ED & NN \\
    \hline
    $\lambda=-0.23,\ N=8, N_s = 24$ & 39.34564427 & 39.24546(13) \\
    \hline
    $\lambda=-0.26,\ N=9, N_s = 27$ & 52.63445418 & 52.17987(51) \\
    \hline
  \end{tabular}
  \caption{Energies (in units of $\hbar^2/m a_0^2$) computed from band projected ED and NN-VMC. All calculations were done for $r_s \approx 3.43$.  }
  \label{tab:NNenergiesvsED}
\end{table}

\begin{figure}[t!]
    \centering
    \includegraphics[width=1.0\linewidth]{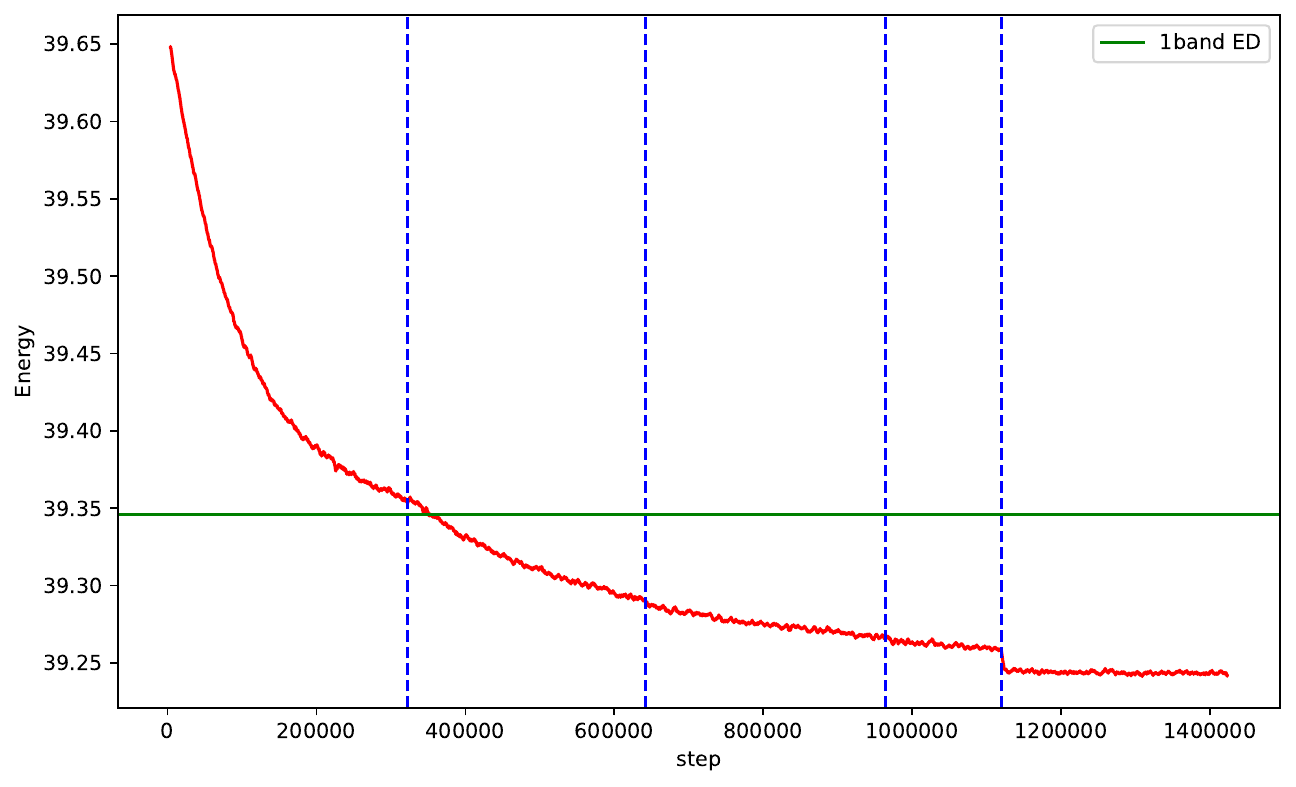}
    \caption{Energy (in units of $\hbar^2/m a_0^2$) as a function of training steps for $N = 8$ , $N_s = 24$ and $\lambda = -0.23$. Energy at a certain step is averaged over the previous 4000 steps. The green line denotes 1 band projected ED. The vertical dashed lines denote instances when the training was stopped and resumed with smaller learning rates.}
    \label{fig:trainingcurve}
\end{figure}

In Fig. \ref{fig:trainingcurve}, we show a training curve obtained for $N = 8$ and $\lambda = -0.23$ where the parameters are initialized randomly which is the case throughout the paper.

\newpage

\bibliography{references}

\end{document}